# DISEÑO E IMPLEMENTACIÓN DE UN GATEWAY IOT MULTIPROTOCOLO

# DESIGN AND IMPLEMENTATION OF A MULTIPROTOCOL IOT GATEWAY

José Macías.* Harold Pinilla.** Wilder Castellanos. *** José David Alvarado.****
Andrés Sánchez. *****

**Resumen**: En los próximos años se espera la interconexión de una gran cantidad de dispositivos lo cual conllevará a una nueva forma de interacción entre el mundo real y el virtual. En este prometedor escenario, conocido como el Internet de las Cosas (IoT), se espera que diferentes objetos, tal como sensores, robots industriales, automóviles, electrodomésticos, entre otros, estén conectados continuamente a Internet. Uno de los principales retos que impone el Internet de las Cosas es la interconexión de dispositivos con características heterogéneas principalmente en términos de las capacidades de comunicación de los dispositivos y los protocolos de red utilizados. Es por esto que, el modelo de interconexión de los diferentes dispositivos involucra un dispositivo intermediario conocido como Gateway. Este gateway sirve como elemento centralizado para la gestión de los dispositivos que conforman una aplicación IoT. Además, resulta esencial para la transmisión de la información a Internet. Esto último se ha convertido en una tarea esencial, principalmente si se tiene en cuenta que muchos dispositivos diseñados para el desarrollo de IoT no contemplan la utilización de las interfaces Ethernet o Wifi, ni la pila de protocolos TCP/IP.

---

* Estudiante Ingeniería Electrónica, Universidad de San Buenaventura, Colombia. Auxiliar de investigación: Universidad de San Buenaventura, Colombia. e-mail:
** Estudiante Ingeniería Electrónica, Universidad de San Buenaventura, Colombia. Auxiliar de investigación: Universidad de San Buenaventura, Colombia. e-mail:
*** Ingeniero Electrónico, Universidad Industrial de Santander, Colombia. Doctor en Telecomunicaciones,
Universitat Politècnica de València, España. Profesor Titular: Universidad de San Buenaventura, Colombia. e-
mail: wcastellanos@usbbog.edu.co
**** Ingeniero Electrónico, Universidad de Cundinamarca, Colombia. Magister en Ingeniería,
Universidad de Ibagué, Colombia. Profesor Asociado: Universidad de San Buenaventura, Colombia. e-
mail: jalvarado@usbbog.edu.co
***** Ingeniero de Sistemas, Universidad Católica de Colombia. Magister en Ingeniería de Sistemas
y Computación, Pontificia Universidad Javeriana, Colombia. Profesor Tiempo Completo
Universidad de San Buenaventura, Colombia. e-mail: aasanchez@usbbog.edu.co

En este artículo se describe la implementación de un gateway IoT que permite el intercambio de datos a través de diferentes tecnologías inalámbricas y el reenvío de dichos datos a Internet. El gateway propuesto incorpora importantes ventajas como: la interconectividad con diferentes protocolos; permite la configuración remota de nodos inalámbricos para la gestión de sensores y actuadores; utiliza un algoritmo flexible para traducir los datos obtenidos por los sensores a un formato uniforme para la transmisión hacia un servidor en la nube; utiliza el protocolo MQTT, el cual permite la difusión de datos de forma simple para minimizar los recursos utilizados del gateway. Con el fin de demostrar el funcionamiento del gateway desarrollado, se implementó una prueba de concepto compuesta por 2 nodos inalámbricos encargados de sensar 6 variables ambientales y transmitiendo dichos datos al gateway a través de diferentes protocolos de comunicación. Los resultados presentados muestran la transmisión de la información de manera simultánea, desde los nodos inalámbricos remotos hacia el gateway. Además, se presentan métricas sobre el consumo de energía en los diferentes dispositivos que componen el escenario desarrollado.

**Palabras clave**: Internet de las Cosas, IoT gateway, Protocolos IoT, Redes Inalámbricas.

**Abstract:** In the coming years, the interconnection of a large number of devices is expected, which will lead to a new form of interaction between the real and the virtual world. In this promising scenario, known as the Internet of Things (IoT), it is expected that different objects, such as sensors, industrial robots, cars, appliances, etc., will be continuously connected to the Internet. One of the main challenges of the Internet of Things is the interoperability of highly heterogeneous devices, mainly in terms of the communication capabilities and network protocols used. As consequence, the interconnection model of the different devices involves an intermediary device, known as gateway. This gateway is a centralized element for the management of the devices that make up an IoT application. In addition, it is essential for the transmission of information to the Internet, especially when many IoT devices are not IP-based.

This paper describes the implementation of an IoT gateway that allows the exchange of data through different wireless technologies and forwarding of such data to the Internet. The proposed gateway has important advantages such as: supporting for multiprotocol interconnectivity; the remote configuration of wireless nodes for sensor and actuators management; a flexible algorithm to translate the data obtained by sensors into a uniform format for transmission to a cloud server; low energy consumption due to efficient data transfer over the MQTT protocol. In order to demonstrate the usefulness of the developed gateway, a proof-of-concept test was implemented. The implemented scenario consists of 2 wireless nodes responsible for sensing environmental variables and transmitting data to the gateway node through different communication protocols. The obtained results show the feasibility for simultaneous data transmission from the remote wireless nodes to the gateway. Metrics on energy consumption in the devices are also presented.

**1. Introducción**
El Internet de las Cosas es un nuevo paradigma tecnológico que promueve la visión de una red global de máquinas y dispositivos capaces de interactuar entre ellos [1], [2]. La perspectiva de las aplicaciones del IoT es que todos estos dispositivos se comuniquen e interactúen entre sí [3], siendo capaces de adquirir información, procesarla y compartirla.

Estimaciones recientes proyectan al Internet de las Cosas (IoT) y al ecosistema asociado con un mercado potencial de USD$7,1 billones en el 2020 [4] y una inversión de alrededor de USD$6 trillones en soluciones IoT por parte de la Industria. Además, se espera que en 2022 el tráfico M2M (Machine-to-Machine) constituya el 45% de todo el tráfico de Internet [5]. Este significativo aumento del tráfico entre dispositivos será debido al importante crecimiento del número de máquina conectadas, el cual de acuerdo con el McKinsey Global Institute ha sido de aproximadamente 300% en los últimos 5 años y se prevé que se pase de una base instalada de 15.400 millones de dispositivos en 2015 a 30.700 millones de dispositivos en 2020 y 75.400 millones en 2025 [6].

Desde el punto de vista de la economía, también será considerable el efecto que supone la incursión del IoT. En este sentido se estima que el impacto anual en la economía global causado por el IoT esté en el rango de 2.7 a 6.2 billones de dólares en el 2025 [7]. Estas previsiones dejan ver un nuevo modo de interacción en el mundo físico, inspirado en la idea de ubicuidad, donde todos los objetos que nos rodean (sensores, automóviles, refrigeradoras, termostatos, robots industriales, tables, smartphones, etc) se puedan conectar a Internet en cualquier momento y en cualquier lugar. Por lo tanto este nuevo modelo se proyecta como una tecnología que lo cambiara todo, incluso a nosotros mismos, razones que ha llevado a la Unión Internacional de las Telecomunicaciones (UIT) ha considerar el IoT como la mayor fuerza tecnológica para los próximos años [8].

Para que dichas previsiones se puedan convertir en realidad, varios retos deben ser superados. Principalmente los relacionados con los temas de seguridad e interoperabilidad de los dispositivos. La interoperabilidad no es un tema menor debido al alto grado de heterogeneidad de los dispositivos disponibles. Esta heterogeneidad se acentúa principalmente si se analizan las capacidades de comunicación (protocolos, tecnologías y hardware) de los elementos que componen un sistema del IoT. Algunas cifras que evidencian la magnitud del problema son proporcionadas por algunos estudios preliminares [9], por ejemplo, recientemente se han identificado alrededor de 10 diferentes estándares de comunicación predominantes en IoT, entre los que se destacan: Bluetooth, WiFi, ZigBee, 6LoWPAN, LoRa y Sig-Fox. Lo anterior supone una dificultad durante la implementación de un sistema real en donde es necesario garantizar una completa interoperabilidad. En particular, el requisito de interoperabilidad es uno de los grandes retos que debe abordarse para la integración y el desarrollo de nuevas plataformas IoT [10]. Algunos estudios, como el desarrollado por Mckinsey Institute, estiman que alrededor del 40 % de los beneficios potenciales de IoT, podrían perder debido a la falta de interoperabilidad entre los diferentes dispositivos [11]. Solucionar el problema de interoperabilidad permitirá eliminar los denominados ecosistemas cerrados [12], obteniendo el verdadero valor de IoT, es decir, datos que se adquieren y se transmiten mediante la interacción entre dispositivos [5].

En este artículo se describe la implementación de un gateway multiprotocolo, capaz de establecer el intercambio de datos con varios nodos inalámbricos, a través una red WiFi, Bluetooth y Zigbee. Esto con el fin de contribuir al desarrollo de sistemas IoT flexibles, escalables y heterogéneos. El gateway, también establece una conexión con un servidor en Internet donde se tiene una herramienta software desarrollada para realizar la analítica de los datos. La evaluación del gateway propuesto, se realizó por medio de un caso de uso, el cual consistió en la implementación de dos nodos inalámbricos, cada uno con 6 sensores. Dichos nodos transmitieron simultáneamente, y a través de diferentes protocolos de comunicación, los datos registrados. Se analizó el rendimiento del gateway a través del registro de las métricas de porcentaje de uso de la CPU y la memoria RAM, así como el análisis del throughput alcanzado en cada interface de red.

## 2. Trabajos Relacionados

El problema de la interoperabilidad de los diferentes dispositivos IoT puede presentarse a diferentes niveles. Por ejemplo, puede presentarse a nivel de los protocolos de interconexión ya que los nodos sensores pueden transmitir por medio de numerosos estándares de comunicación, tal como ZigBee, WiFi, Bluetooth y LoRA. También puede existir problema de interoperabilidad a nivel del formato de los datos y debido a las incompatibilidades entre dispositivos. Estas características heterogéneas a diferentes niveles se deben a la amplia variedad de tecnologías desarrolladas para la construcción de aplicaciones de IoT. Esto se puede comprobar en las referencias [9], [10]

Algunos gateways han sido propuestos para mejorar la interconectividad entre los dispositivos. Por ejemplo, en la referencia [13] se describe la implementación de un gateway IoT dedicado al monitoreo y control remoto de una piscina. Este fue desarrollado mediante una tarjeta Raspberry Pi. El gateway permite la comunicación bidireccional y el intercambio de datos entre el usuario y la red de sensores, lo cuales fueron implementados con Arduino.

El sistema desarrollado tiene sensores de temperatura, humedad, luminosidad y nivel de agua. Sin embargo, todas las comunicaciones entre la red de sensores y el gateway se realiza por medio de un cable USB, por lo cual se hace compleja su implementación en un ambiente real, debido a la dependencia de una conexión cableada. Otro gateway es el propuesto en [14], el cual está basado en la plataforma IoTivity [15]. Este gateway está centrado en garantizar la interoperabilidad entre dispositivos que no tienen una comunicación basada en protocolo IP. Para cumplir con este objetivo, tiene implementado el protocolo CoAP (Constrained Application Protocol) [16]. Una de las ventajas de este gateway es la opción de autoconfiguración, lo cual permite que al conectarse un nuevo nodo sensor, este establezca un diálogo con el gateway con el fin de ser reconocido como nuevo integrante del sistema y la configuración de los parámetros básicos de operación. También Khanchuea et al [17] desarrollaron un gateway IoT. Este gateway está desarrollado con la tarjeta ESP32. Con esta plataforma se establece una red WiFi tipo mesh para intercambiar información con una red de sensores. Los nodos sensores utilizan el sensor DHT22 para registrar valores de temperatura y humedad. Estos nodos establecen una red ZigBee para enviar los datos sensados a un nodo enrutador. Este nodo enrutador toma los datos recibidos por ZigBee y los retransmite por WiFi al nodo gateway. Aunque en este sistema hay una co-existencia de dos redes inalámbricas (una ZigBee y otra WiFi), en realidad el nodo gateway solamente establece conexiones WiFi, por lo que carece de la propiedad de ser multiprotocolo. Otras otros trabajos relacionados con el desarrollo de gateways para IoT, son los presentados en las referencias [18] [19]. Dichos gateways tienen como función principal la conversión de protocolos principalmente entre protocolos Zigbee, Bluetooth, WiFi y Ethernet. En ambos casos los datos son almacenados y visualizados, por medio de un servidor web embebido dentro del mismo gateway, lo cual no es del todo práctico si se analiza la posibilidad de una aplicación real. Lo más idóneo sería la utilización de un servicio de alojamiento en la nube.

### 3. Marco Teórico: El gateway en aplicaciones del IoT

En términos generales un gateway o puerta de enlace, es un dispositivo que permite la comunicación entre redes con diferentes protocolos. Para esto es necesario traducir los formatos de los datos recibidos al protocolo de destino. Para el caso de IoT, un gateway es un dispositivo que soporta comunicaciones con una variedad de protocolos de comunicación y formatos de datos. Esto permite la interconexión de múltiples tipos de sensores y la agregación de los datos provenientes de estos para posteriormente ser enviados a otro segmento de la red o a Internet. Este es el principal objetivo de un gateway, servir de puente entre varios dominios de red con una red pública o Internet, resolviendo el problema de la heterogeneidad entre estos dominios [20]. Al mismo tiempo, el gateway se convierte en un elemento idóneo para ejecutar funciones de control de la red, ya que mientras intercambia mensajes con los nodos sensores, puede mapear la red y establecer un conocimiento completo de la red.

Los diferentes tipos de gateways para IoT pueden ser clasificados en pasivos, semi-automáticos y completamente automáticos. Los gateways pasivos son aquellos que necesitan que los nuevos dispositivos IoT que se incorporen a la red, deben ser configurados manualmente. De igual manera deben ser retirados los antiguos dispositivos que ya no formen parte de la red. Un ejemplo de este tipo gateways es el presentado en la referencia [21]. Los gateways semi-automáticos pueden generar un enlace con el nuevo dispositivo pero no pueden soportar automáticamente el establecimiento de todos los parámetros de configuración [22]. Finalmente, los gateways que son completamente automáticos permiten la autoconfiguración de nuevos dispositivos IoT y de esta manera resolver rápidamente problemas de heterogeneidad durante la transmisión de los datos [23].

## 4. Gateway multiprotocolo propuesto

A continuación, se hará una descripción de la arquitectura y los componentes hardware y software del gateway IoT desarrollado. Posteriormente se describirán los diferentes componentes que integran el gateway y sus principales funciones.

### 4.1. Arquitectura del sistema

Los principales componentes de la arquitectura del gateway se resumen en la Figura 6. De esta figura se identifican cinco principales funciones en el gateway IoT: *(i)* interconexión multiprotocolo con nodos remotos, *(ii)* transformación de datos, *(iii)* conversión de protocolos *(iv)* comunicación con la nube y *(v)* Interfaz con el usuario.

- Interconexión: la arquitectura propuesta permite la interconexión de varios nodos inalámbricos intercambiando información, de manera simultánea, a través de diferentes tipos de protocolos (tal como ZigBee, WiFi y Bluetooth). Esta comunicación es bidireccional, ya que el gateway recoge los datos registrados por los nodos inalámbricos a través de sus sensores, pero además, puede enviarse información a los nodos para operar sobre los dispositivos actuadores.

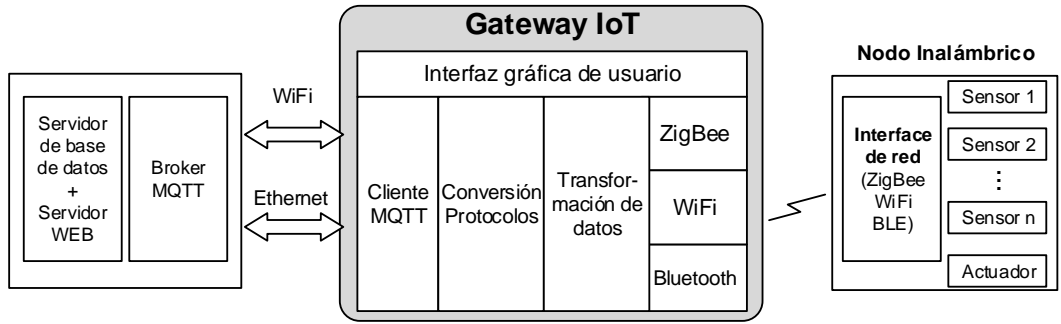

**Figura 6. Arquitectura del gateway IoT desarrollado**

- Transformación de datos: consiste en la normalización y el establecimiento de un formato estándar para, posteriormente, enviar estos datos al servidor de base de datos. Esta transformación es indispensable ya que los datos provienen de nodos heterogéneos que realizaron la captura diferentes tipos de sensores y que llegan a través de diferentes protocolos de red. Para estandarizar la representación de los datos se eligió el formato JSON, el cual tiene importantes ventajas como la sencillez y el bajo consumo de recursos. Al estandarizar el formato para los datos, también se establece un modelo para la identificación de los dispositivos que componen un escenario IoT. En la Figura 7 se muestra un ejemplo de la representación de los datos registrados por el sensor de temperatura y humedad, en formato JSON. Como se puede observar en dicha figura, los primeros tres datos registrados son "node-id", "gps" y "protocol", los cuales corresponden a la identificación del nodo remoto, sus coordenadas y el protocolo por el cual transmitió los datos. Los siguientes datos son *"date"*, *"sensor-id"*, *"value"* y *"magnitude"*, los cuales registran la información del sensor y su lectura. Finalmente están los datos de "gate-id" y "network-id", qe serán utilizados más adelante para identificar el gateway y la red de comunicación. Estos últimos registros serán útiles cuando se tenga un escenario con más de un gateway y/o más de una red de transmisión.

187

```
{
    "node-id": "nodo2",
    "gps": "-",
    "protocol": "wifi",
    "date": "09/13/19-08:59:18",
    "sensor-id": "Temperature",
    "value": "19.9"
    "magnitude": "celcius",
    "gate-id": "-",
    "network-id": "-",
}
{
    "node-id": "nodo2",
    "gps": "-",
    "protocol": "wifi",
    "date": "09/13/19-08:59:18",
    "sensor-id": "Humidity",
    "value": "48.2"
    "magnitude": "percent",
    "gate-id": "-",
    "network-id": "-",
}
```

**Figura 7. Ejemplo del formato estandarizado para la transmisión de los datos**

- Conversión de protocolos: el gateway IoT propuesto, actúa como un puente o traductor entre diferentes protocolos, principalmente entre los protocolos ZigBee, Bluetooth, WiFi y Ethernet. De modo que es necesario que el gateway esté continuamente escuchando peticiones de conexión desde sus interfaces de red. La recepción de datos por cualquier protocolo de comunicación involucra la extracción de la carga útil y la correspondiente estructuración de un nuevo mensaje con el formato estandarizado. Cuando los datos viajan con destino a la base de datos para su almacenamiento y posterior visualización, el protocolo utilizado es MQTT (Message Queuing Telemetry Transport)[24]. Este protocolo es recomendado en escenarios de red en los cuales el consumo de ancho de banda debe ser reducido y donde los dispositivos involucrados en la comunicación tienen baja capacidad de procesamiento y de memoria. El protocolo MQTT funciona bajo un modelo publicación – suscripción que utiliza tres componentes: un *broker*, un *suscriptor* y un *publicador*. Un dispositivo se puede registrar como *suscriptor* a un *tópico* de interés con el fin de obtener información publicada en ese *tópico*. El *publicador* es el generador de los datos para un *tópico* específico. Los datos de un *tópico* son transmitidos al *suscriptor* por intermedio del *Broker.* De ahí que un *Broker* se puede considerar como un servidor que enruta los mensajes publicados a los subscriptores.

- Comunicación con la nube: consiste en recibir los datos provenientes de los sensores ya formateados y enviarlos a un bróker MQTT alojado en Internet (iot.eclipse.org). También se encarga de recibir los parámetros de configuración que un usuario puede ingresar a través de la interfaz gráfica y transmitirlos a los nodos remotos, también usando el protocolo MQTT. En el primer caso, comunicación del gateway hacia el bróker, el gateway utiliza la librería Eclipse Paho Python Client y actúa como publicador. En el segundo caso, es decir durante para la transmisión desde el gateway a los nodos inalámbricos, el gateway opera como un publicador y los nodos inalámbricos como subscriptores. Debido a que los nodos no cuentan con conexión a Internet, es necesario también que el gateway tenga instalado un bróker para este tipo de intercambio de mensajes. El bróker instalado en el gateway fue el conocido bróker Mosquitto [25], el cual debido a sus características de bajo consumo de recursos y reducida información de sobrecarga, es idóneo para ser instalado en tarjetas embebidas.

- Interfaz con el usuario: finalmente, la quinta función del gateway es el registro de algunos parámetros de configuración del sistema, a través de una interfaz gráfica. Dicha interfaz gráfica permite que el usuario pueda registrar nuevos nodos y/o sensores, la modificación de la frecuencia de las lecturas, la asignación de protocolos de comunicación, la visualización de las medidas registradas en la base de datos, la creación de reglas para anuncios y alarmas, entre otras funciones.

Es importante aclarar que, en el modelo planteado, la base de datos donde se almacenan los datos finalmente y la plataforma web donde se tiene una interfaz de visualización de los datos, están alojados fuera del gateway. Concretamente en un servidor privado en Internet

## 4.2. Hardware utilizado

Para el desarrollo del gateway fue utilizado el kit de desarrollo Samsung Artik 1020, el cual es una tarjeta embebida de alto rendimiento y multiprotocolo que cuenta con la posibilidad de comunicarse inalámbricamente mediante Bluetooth, ZigBee y WiFi. Además cuenta con múltiples puertos de I/O capaz comunicarse mediante los módulos I2C, SPI, UART, entre otros. Un resumen con las principales especificaciones técnicas se muestra en la Figura 8.

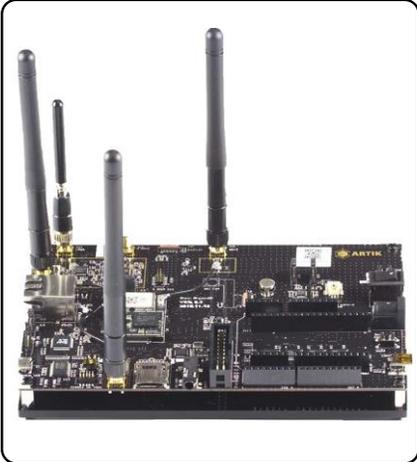

**Figura 8. Principales especificaciones técnicas de la tarjeta Artik 1020**

## 5. Escenario de aplicación del gateway IoT desarrollado

Con el propósito de evaluar y poner en operación el gateway desarrollado, se construyó un caso de uso compuesto por dos nodos inalámbricos remotos que tienen conectados 6 sensores registrando las siguientes variables: temperatura del ambiente, humedad relativa, radiación solar, velocidad del viento, nivel de precipitaciones, dirección del viento. Los nodos inalámbricos transmiten los datos registrados por los sensores al gateway a través de ZigBee, WiFi y Bluetooth. Esta transmisión de datos fue configurada de la siguiente manera: las variables de dirección y velocidad del viento fueron transmitidas utilizando Bluetooth; las variables de radiación y nivel de precipitación por medio de ZigBee; y finalmente, las variables de temperatura y humedad por WiFi.

Un diagrama esquemático que resume el caso de uso implementado, se muestra en la Figura 9.

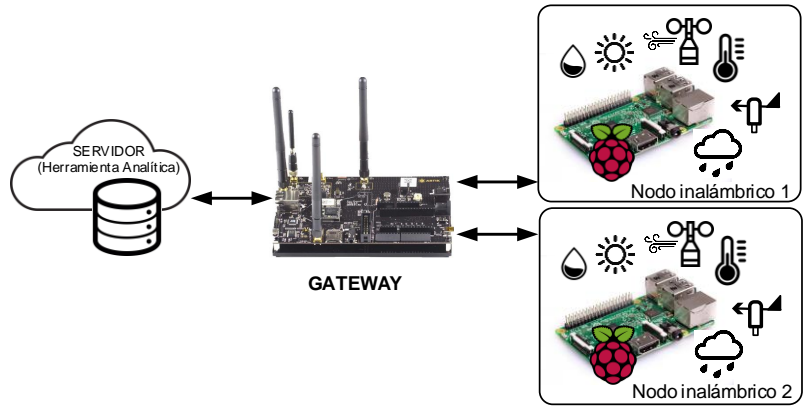

**Figura 9. Caso de uso implementado**

Los nodos inalámbricos fueron implementados con los siguientes componentes hardware y software. La tarjeta base fue una Raspberry Pi 3 Modelo B a la cual estaban conectados los diferentes sensores. La tarjeta fue configurada para ser operada desde el sistema operativo Ubuntu Mate. Los sensores utilizados fueron los siguientes. El sensor utilizado para la medición de la temperatura ambiente y humedad relativa es el AM2315. El sensor Davis 6450 es el encargado de realizar la medición de la radiación solar. Y por último se utilizó el kit SEN-08942 que está conformado por un anemómetro, un pluviómetro y una veleta para determinar la dirección del viento. Un resumen con las principales características de los nodos inalámbricos se muestra en la Figura 10.

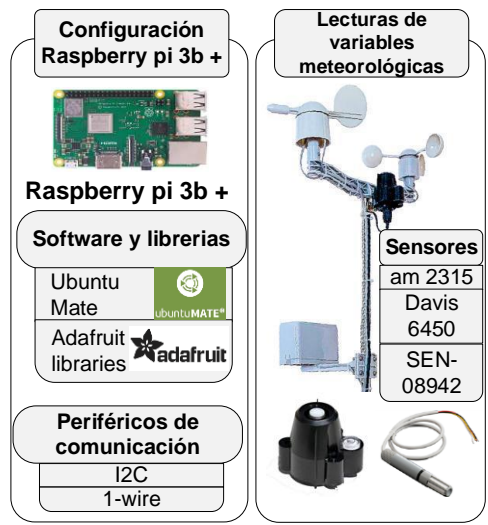

**Figura 10. Principales características de los nodos inalámbricos**

La herramienta de analítica fue desarrollada para soportar arquitecturas de comunicación de tipo *cliente/servidor* y *publicador/suscriptor*, con el gateway. Tiene una arquitectura por capas desacopladas, con Web Service RESTful para la comunicación entre el Backend y Frontend (ver Figura 11). El Backend tiene conexión a bases de datos MySQL 5.7 y MongoDB 4.2 mediante DAO y JPA, desarrollado en JavaEE 7 sobre un servidor Glassfish 4.1. Para la conexión con el *broker* se usa PAHO Eclipse 1.4 y publica un Web Service con seguridad OAuth2 para ser consumido por el Frontend. El Frontend consume el servicio con OAuth2 Client y un token de seguridad, la información la presenta con Bootstrap chart de acuerdo a los datos y variables establecidas.

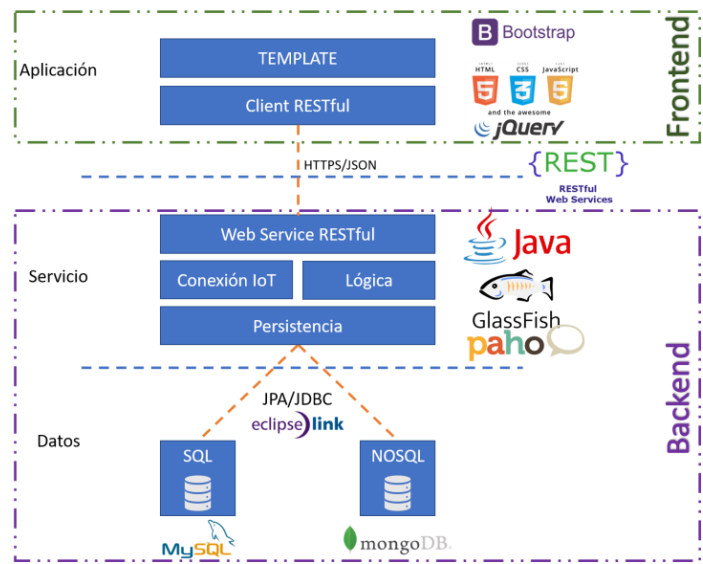

**Figura 11. Arquitectura de la herramienta de analítica**

## 6. Evaluación y Resultados

El desempeño del fue evaluado mediante el registro del tráfico recibido desde los dos nodos inalámbricos a través de los diferentes protocolos. El sistema implementado, realizó la captura de los datos y la transmisión hacia el gateway durante un tiempo de 8 minutos. En la Figura 12 y Figura 13, se muestra la variación del throughput instantáneo del tráfico recibido por las interfaces WiFi y Bluetooth, respectivamente, para cada uno de los dos nodos. Para el tráfico WiFi, se observa que los flujos tienen máximos de tráfico de 8 Kbps. En promedio, el throughput del tráfico del nodo 1 es de 1.25 Kbps y para el nodo 2 de 1.5 Kbps. Estos valores son muy bajos, a pesar de que la frecuencia de transmisión de las variables registradas fue de 6 segundos. Con este bajo nivel de tráfico, se puede afirmar que aún queda margen suficiente en la capacidad del gateway para permitir la interconexión de varios nodos remotos.

En cuanto al tráfico por Bluetooth, el nodo 1 tuvo un throughput medio de 0.95Kbps y el nodo 2 aproximadamente de 0.85Kbps. Esto equivale a aproximadamente 1.8Kbps de media en el tráfico agregado. Tomando como base estos valores, con el tráfico Bluetooth, al igual que con el tráfico WiFi, muestran que se podrían adicionar más nodos transmitiendo al gateway por este protocolo.
De igual forma, en la Figura 14, se muestra el flujo de tráfico que llega al gateway a través del protocolo Zigbee. En la gráfica se muestra el tráfico agregado (nodo 1 + nodo2) y para este caso, se obtuvo un throughput de 16 Kbps.

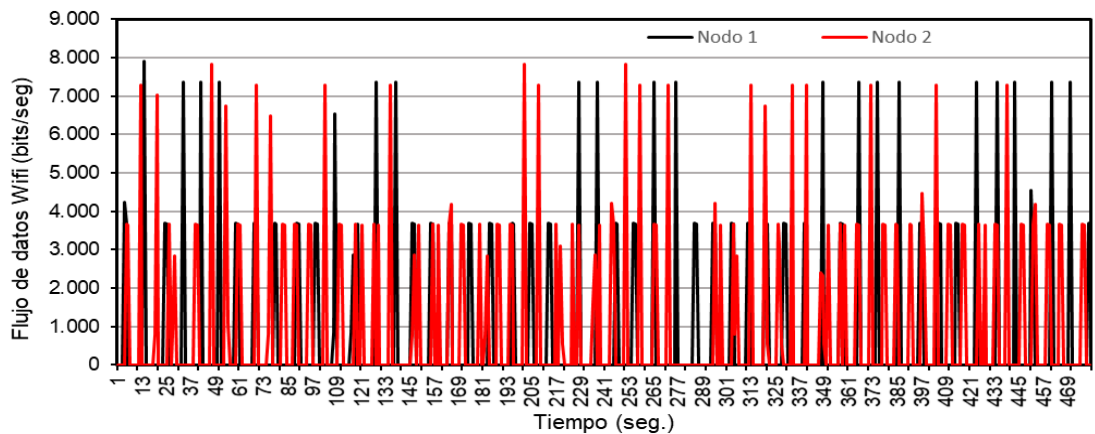

**Figura 12. Tráfico recibido a través de la interface WiFi**

Finalmente, se registraron tanto el porcentaje de uso de la CPU, como la memoria libre en la tarjeta embebida en la que se implementó el gateway (ver Figura 15). Las lecturas de estas medidas de rendimiento se tomaron cada 10 segundos durante la operación del sistema. Es decir, durante los 8 minutos en los cuales los nodos estaban transmitiendo los valores registrados por los sensores hacia el gateway. Al analizar el uso de la CPU, se observa que en ningún momento se superó el 5% de uso. Y en cuanto a la memoria RAM, se obtuvo en promedio 1.5 GB de memoria disponible, esto representa una disponibilidad equivalente al 75% del total de memoria instalada en la tarjeta embebida. Los resultados obtenidos indican que los procesos implementados en el gateway tienen un bajo consumo de recursos, lo cual permitirían continuar con la incorporación de más algoritmos y de más nodos remotos haciendo uso del gateway.

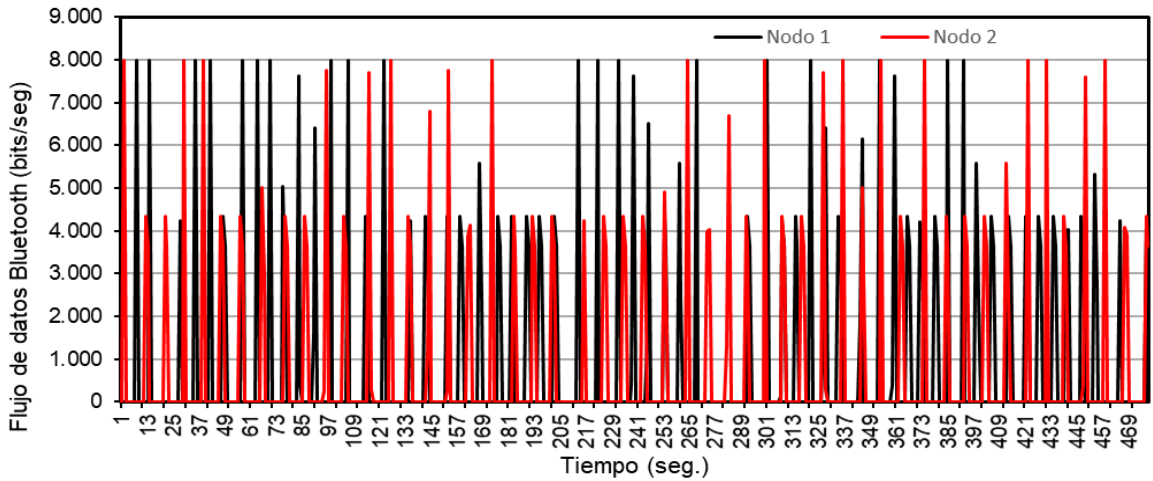

**Figura 13. Tráfico recibido a través de Bluetooth**

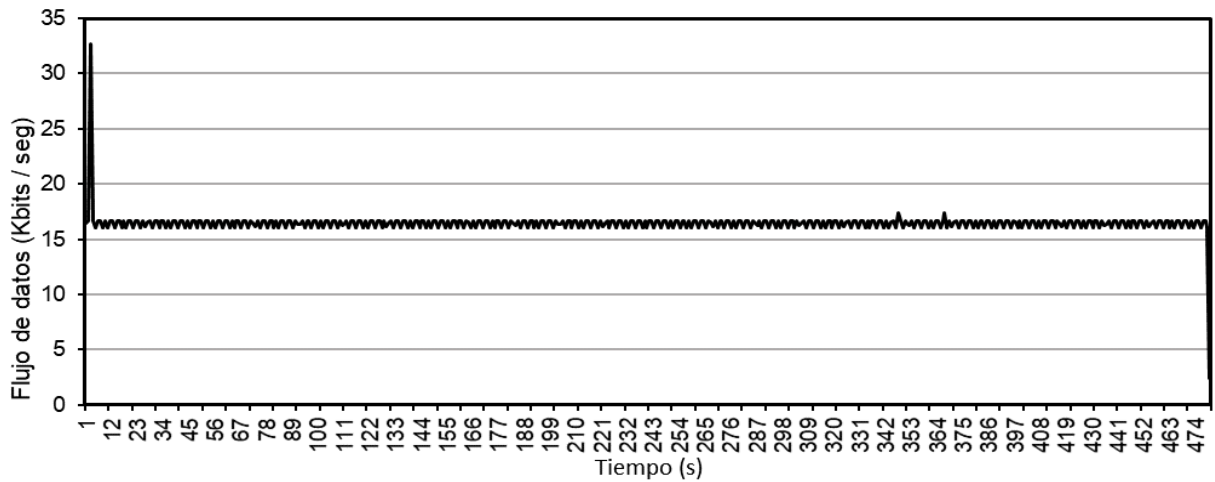

**Figura 14. Tráfico recibido a través de ZigBee**

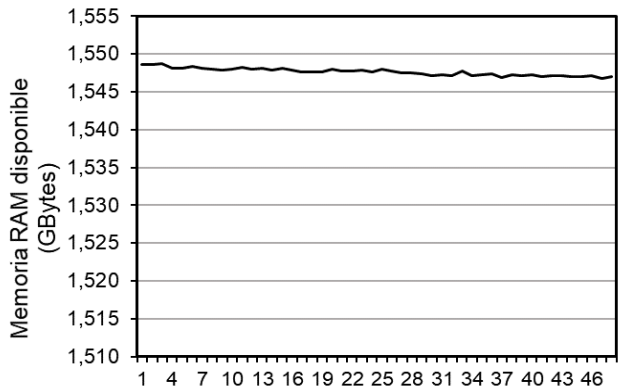
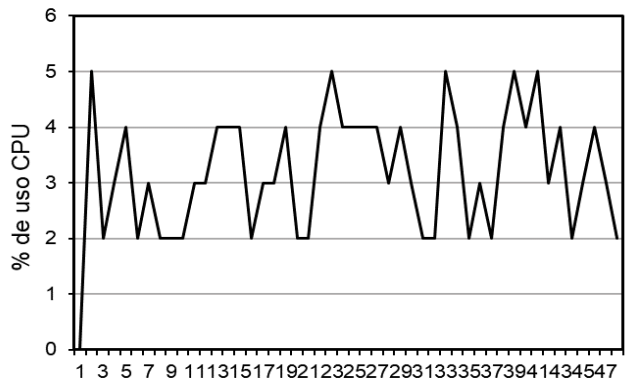

(a) (b)

**Figura 15. Medidas de desempeño del gateway: (a) Memoria RAM disponible y (b) Porcentaje de uso de la CPU**

## 7. Conclusiones

El gateway IoT desarrollado en este trabajo, contribuye a solucionar el problema de interoperabilidad existente entre los dispositivos IoT, principalmente a nivel de interconectividad y del formato de los datos. El gateway implementado actúa como el elemento central en un modelo en el cual varios nodos inalámbricos pueden enviarle datos a través de diferentes protocolos de comunicación, tales como: WiFi, Bluetooth, ZigBee y Ethernet. La evaluación del rendimiento del gateway y la transmisión de datos hacia este, generaron resultados aceptables en términos de consumo de ancho de bando y porcentaje de uso de la CPU y la RAM.

Una de las principales ventajas del gateway propuesto, es que permite la configuración de los nodos inalámbricos remotos y la transmisión de los datos a una herramienta de analítica de datos alojada en un servidor de Internet. Esto brinda flexibilidad en el almacenamiento y visualización de los datos.

Se tiene previsto como trabajo futuro, aumentar el número de nodos inalámbricos que transmiten directamente al gateway, así como la construcción de un caso de estudio en el cual dicho sistema sea aplicado a un escenario de agricultura inteligente, en el cual se cuente, además de los sensores, con un actuador y con autonomía desde el punto de vista energético, a través de energía fotovoltaica.

**Referencias**

[1]    J. Gubbi, R. Buyya, S. Marusic, y M. Palaniswami, «Internet of Things (IoT): A vision, architectural elements, and future directions», *Future Gener. Comput. Syst.*, vol. 29, n.º 7, pp. 1645–1660, sep. 2013.

[2]    I. Lee y K. Lee, «The Internet of Things (IoT): Applications, investments, and challenges for enterprises», *Bus. Horiz.*, vol. 58, n.º 4, pp. 431–440, jul. 2015.

[3]    D. Miorandi, S. Sicari, F. De Pellegrini, y I. Chlamtac, «Internet of things: Vision, applications and research challenges», *Ad Hoc Netw.*, vol. 10, n.º 7, pp. 1497-1516, sep. 2012.

[4]    D. Lund, C. MacGillivray, V. T.-\ldots D. C. (IDC) \ldots, y undefined 2014, «Worldwide and regional internet of things (iot) 2014–2020 forecast: A virtuous circle of proven value and demand», *business.att.com*.

[5]    D. Evans, «The internet of things: How the next evolution of the internet is changing everything», 2011.

[6]    S. Lucero, «IoT platforms: enabling the Internet of Things», 2016.

[7]     J. Manyika, M. Chui, J. Bughin, R. Dobbs, P. Bisson, y Marrs, «Disruptive technologies: Advances that will transform life, business, and the global economy», 2013.

[8]     P. Strategy, I. T. U., & Unit, «ITU Internet Reports 2005: The internet of things», Geneve, 2005.

[9]     J. Alvarado, L. Luis, W. Castellanos, y A. Barrera, «Embedded systems for Internet of Things (IoT) applications: a comparative study», en *2018 Congreso Internacional de Innovacion y Tendencias en Ingenieria (CONIITI)*, Bogotá, 2018, p. 6.

[10]    A. Al-Fuqaha, M. Guizani, M. Mohammadi, M. Aledhari, y M. Ayyash, «Internet of Things: A Survey on Enabling Technologies, Protocols, and Applications», *IEEE Commun. Surv. Tutor.*, vol. 17, n.º 4, pp. 2347–2376, 2015.

[11]    J. Manyika *et al.*, «The Internet of Things: Mapping the value beyond the hype», McKinsey Global Institute, June, 2015.

[12]    P. Desai, A. Sheth, P. A.-M. S. (MS), U. 2015, y U. 2015, «Semantic gateway as a service architecture for iot interoperability», en *2015 IEEE International Conference on Mobile Services*, New York, NY, USA, 2015.

[13]    A. Glória, F. Cercas, y N. Souto, «Design and implementation of an IoT gateway to create smart environments», *Procedia Comput. Sci.*, vol. 109, pp. 568-575, ene. 2017.

[14]    B. Kang y H. Choo, «An experimental study of a reliable IoT gateway», *ICT Express*, vol. 4, n.º 3, pp. 130-133, sep. 2018.

[15]    Linux Foundation, «IoTivity», 2019. [En línea]. Disponible en: https://iotivity.org/. [Accedido: 11-sep-2019].

[16]    Z. Shelby, K. Hartke, y C. Bormann, «The Constrained Application Protocol (CoAP), RFC 7252». Internet Engineering Task Force (IETF) Standard, 2014.

[17]    K. Khanchuea y R. Siripokarpirom, «A Multi-Protocol IoT Gateway and WiFi/BLE Sensor Nodes for Smart Home and Building Automation: Design and Implementation», en *2019 10th International Conference of Information and Communication Technology for Embedded Systems (IC-ICTES)*, 2019, pp. 1-6.

[18]  S. Guoqiang, C. Yanming, Z. Chao, y Z. Yanxu, «Design and Implementation of a Smart IoT Gateway», en *2013 IEEE International Conference on Green Computing and Communications and IEEE Internet of Things and IEEE Cyber, Physical and Social Computing*, 2013, pp. 720-723.

[19]  D. C. Yacchirema Vargas y C. E. Palau Salvador, «Smart IoT Gateway For Heterogeneous Devices Interoperability», *IEEE Lat. Am. Trans.*, vol. 14, n.º 8, pp. 3900-3906, ago. 2016.

[20]  T. Zachariah, N. Klugman, B. Campbell, J. Adkins, N. Jackson, y P. Dutta, «The Internet of Things Has a Gateway Problem», en *Proceedings of the 16th International Workshop on Mobile Computing Systems and Applications - HotMobile '15*, Santa Fe, New Mexico, USA, 2015, pp. 27-32.

[21]  K. A. Emara, M. Abdeen, y M. Hashem, «A gateway-based framework for transparent interconnection between WSN and IP network», en *IEEE EUROCON 2009*, 2009, pp. 1775–1780.

[22]  L. Wu, Y. Xu, C. Xu, y F. Wang, «Plug-configure-Play Service-oriented Gateway - For Fast and Easy Sensor Network Application Development», presentado en 2nd International Conference on Sensor Networks, 2013, pp. 53-58.

[23]  B. Kang, D. Kim, y H. Choo, «Internet of Everything: A Large-Scale Autonomic IoT Gateway», *IEEE Trans. Multi-Scale Comput. Syst.*, vol. 3, pp. 206-214, jul. 2017.

[24]  «SO/IEC 20922:2016 Message Queuing Telemetry Transport (MQTT)». ISO standard, 2016.

[25]  R. A. Light, «Mosquitto: server and client implementation of the MQTT protocol.», *J. Open Source Softw.*, vol. 2, n.º 13, p. 265, 2017.